\documentclass[superscriptaddress,secnumarabic,
amssymb,amsmath,nobibnotes,aps,prd,showkeys,showpacs,nofootinbib]{revtex4}%
\usepackage{graphicx}
\usepackage{epsf}
\usepackage{bm}
\usepackage{amsmath}
\usepackage{amsfonts}
\usepackage{amssymb}
\usepackage{epstopdf}
\usepackage{natbib}%
\providecommand{\U}[1]{\protect\rule{.1in}{.1in}}
\newcommand{\be}{\begin{equation}}
\newcommand{\ee}{\end{equation}}

\newcommand{\mincir}{\raise
-3.truept\hbox{\rlap{\hbox{$\sim$}}\raise4.truept\hbox{$<$}\ }}
\newcommand{\magcir}{\raise
-3.truept\hbox{\rlap{\hbox{$\sim$}}\raise4.truept\hbox{$>$}\ }}

\begin{document}
\title{Will there be future deceleration? A study of particle creation mechanism in non-equilibrium thermodynamics}
\author{Supriya Pan}
\email{span@research.jdvu.ac.in}
\affiliation{Department of Mathematics, Jadavpur University, Kolkata-700032, West Bengal, India}
\author{Subenoy Chakraborty}
\email{schakraborty@math.jdvu.ac.in}
\affiliation{Department of Mathematics, Jadavpur University, Kolkata-700032, West Bengal, India}

\begin{abstract}
The paper deals with non-equilibrium thermodynamics based on adiabatic particle creation mechanism with the motivation of considering it as an alternative choice to explain the recent observed accelerating phase of the universe. Using Friedmann equations, it is shown that the deceleration parameter ($q$) can be obtained from the knowledge of the particle production rate ($\Gamma$). Motivated from thermodynamical point of view, cosmological solutions are evaluated for the particle creation rates in three cosmic phases, namely, inflation, matter dominated and present late time acceleration. The deceleration parameter ($q$) is expressed as a function of the redshift parameter ($z$), and its variation is presented graphically. Also, statefinder analysis has been presented graphically in three different phases of the universe. Finally, two non-interacting fluids with different particle creation rates are considered as cosmic substratum, and deceleration parameter ($q$) is evaluated. It is examined whether more than one transition of $q$ is possible or not by graphical representations. \\

\end{abstract}
\maketitle
\section{Introduction}
There was a dramatic change in our knowledge of the evolution history of the universe based on the standard cosmology at the end of the last century due to some observational predictions from type Ia Supernova \cite{Riess1, Perlmutter1} and others \cite{Komatsu1, Sanchez1}. Riess et al. \cite{Riess1} and Perlmutter et al. \cite{Perlmutter1} observed that distant Supernovae at redshift $z \sim 0.5$, and $\Delta m \sim 0.25$ mag are found to be about 25\% fainter than the prediction from standard cosmology, and hence, they concluded that the universe at present is undergoing through an accelerated expansion rather than deceleration (as predicted by the standard cosmology). This present accelerating phase was also supported by the Cosmic Microwave Background (CMB) \cite{Komatsu1}, and the Baryon Acoustic Oscillations (BAO) \cite{Sanchez1}. The explanation of this unexpected accelerating phase is a great challenge to the theoretical physics. Since the discovery of this accelerating universe, people are trying to explain this observational fact in two different ways-- either modifying the Einstein's gravity itself, or, by introducing some unknown kind of matter in the framework of Einstein gravity. In the second option, cosmological constant is the common choice for this unknown matter. But it suffers from two serious problems-- the measured value of the cosmological constant is far below the prediction from quantum field theory, and secondly, the coincidence problem \cite{Carroll1}. So, people choose this unknown matter as some kind of dynamical fluid with negative and time dependent equation of state, and is termed as dark energy (DE). Though a lot of works have been done with several models of DE (see references \cite{Padmanabhan1, Peebles1, Copeland1, Yoo1} for reviews) but still its origin is totally mysterious to us.\\

Among other possibilities to explain the present accelerating stage, inclusion of back reaction in the Einstein's field equations through an effective ($-$ve) pressure is much relevant in the context of cosmology, and the gravitational production of particles [radiation, or cold dark matter (CDM)] provides a mechanism for cosmic acceleration \cite{Prigogine1, Calvao1, Lima1, Lima2}. In particular, in comparison with dark energy models, the particle creation scenario has a strong physical basis: the non-equilibrium thermodynamics. Also, the particle creation mechanism not only unifies the dark sectors (DE+ DM) \cite{Lima2}, but also it contains only one free parameter as we need only a single dark component (DM). Further, Statistical Bayesian analysis with one free parameter should be preferred along with the hierarchy of cosmological models \cite{Guimaraes1}. So, the present particle creation model which simultaneously fits the observational data and alleviates the coincidence and fine-tuning problems, is better compared to the known (one-parameter) models namely (i) the concordance $\Lambda$CDM which however suffers from the coincidence and fine-tuning problems \cite{Zlatev1, delCampo1, Steinhardt1} and (ii) the brane world cosmology \cite{Deffayet1} which does not fit the SNIa+ BAO+ CMB (shift-parameter) data \cite{Basilakos1}. Furthermore, it should be mentioned that the thermodynamics of dark energy has been studied in equilibrium and non-equilibrium situations in the literature \cite{Jamil1, Jamil2, Jamil3, Jamil4}.\\

The homogeneous and isotropic flat FLRW model of the universe is chosen as an open thermodynamics system which is adiabatic in nature. Although the entropy per particle is constant for this system, still there is entropy production due to expansion of the universe (i.e., enlargement of the phase space) \cite{Zimdahl1}. As a result, the dissipative pressure (i.e., bulk viscous pressure) is linearly related to the particle creation rate $\Gamma$ \cite{Zimdahl1, Chakraborty1}. Further, using the Friedmann's equations, one can relate $\Gamma$ to the evolution of the universe [see equation (\ref{eqn8}) below]. Choosing $\Gamma$ as a function of the Hubble parameter from thermodynamical view point, it is possible to describe different phases of the evolution of the universe and $q$ can be obtained as a function of $z$, the red shift parameter. Finally, we consider two components of matter which have different particle creation rate \cite{Harko1}, and $q$ has been evaluated and plotted to examine whether more than one transition of $q$ is possible or not. In formulation of the general theory of relativity, in terms of the spin connection coefficients, it has been shown \cite{Arbuzov2010} that the cosmological evolution of the metrics is induced by the dilaton without the inflation hypothesis and the $\Lambda$ term. Further, it is found that the dilaton evolution yields the vacuum creation of matter, and the dilaton vacuum energy plays a role of the dark energy. On the other hand, in the Hamiltonian approach to the gravitational model with the aid of Dirac-ADM foliation, Pervushin et al. \cite{Pervushin2012} showed a natural separation of the dilatonic and gravitational dynamics in terms of the Maurer-Carton forms. As a result, the dominance of the Casimir vacuum energy of physical fields provides a good description of the type Ia Supernovae luminosity distance-redshift relation. Furthermore, introducing the uncertainty principle at the Planck's epoch, it is found that the hierarchy of the universe's energy scales is supported by the observational data. Also, this Hamiltonian dynamics of the model describes the effect of an intensive vacuum creation of gravitons and the minimal coupling scalar (Higgs) bosons in the early universe.\\

Moreover, the motivation of the present work in the framework of the particle creation mechanism comes from some recent related works. It has been shown in Refs. \cite{SSS2014, SS2014}, that the entire cosmic evolution from inflationary stage can be described by particle creation mechanism with some specific choices of the particle creation rates. As these works show late-time acceleration without any concept of dark energy, so, it is very interesting to think of the particle creation mechanism as an alternative way of explaining the idea of dark energy. The present work is an extension of these works by considering two fluid system as the cosmic fluid. The paper is organized as follows: section II deals with non-equilibrium thermodynamics in the background of particle creation mechanics while several choices of $\Gamma$ as a function of the  Hubble parameter are shown in section III and the deceleration parameter $q$ has been presented both analytically and graphically. A field theoretic analysis of the particle creation mechanism is presented in section IV. Section IV is related to interacting two dark fluids having different particle creation rates and it is examined whether two transitions for $q$ are possible or not. Finally, there is a summary of the work in section VI.
\section{Non equilibrium thermodynamics: Mechanism of particle creation}

Suppose the homogeneous and isotropic flat FLRW model of the universe is chosen as an open thermodynamical system. The metric ansatz takes the form

\begin{equation}
ds^2= -dt^2+a^2(t)\left[dr^2+r^2 (d \theta ^2+ \sin^2 \theta d \phi ^2)\right].\label{flrw-eqn1}
\end{equation}
Then the Friedmann equations are

\begin{equation}
3H^2= \kappa \rho,~~~~~~~~\mbox{and},~~~~~~~2 \dot{H}= -\kappa (\rho+p+\Pi),\label{friedmann-eqn2}
\end{equation}
where $\kappa= 8 \pi G$, $H= \dot{a}/a$ is the Hubble rate, $a= a (t)$ is the scale factor of the universe, $\rho$ and $p$ are the total energy density and the thermodynamical pressure of the cosmic fluid, $\Pi$ is related to some dissipative phenomena (bulk viscous pressure), and the overdot denotes the derivative with respect to the cosmic time $t$. It should be noted that there are several choices of $\Pi$ and corresponding solutions in the literature \cite{Barrow1}. The energy conservation relation reads

\begin{equation}
\dot{\rho}+3H(\rho+p+\Pi)= 0.\label{conservation-eqn3}
\end{equation}
As the particle number is not conserved (i.e., $N^{\mu} _{;\mu}$ $\neq$ 0), so, the modified particle number conservation equation takes the form \cite{Harko1}

\begin{equation}
\dot{n}+\Theta n= n \Gamma,\label{eqn4}
\end{equation}
where $n= N/V$ is the particle number density, $N$ is the total number of particles in a co-moving volume $V$, $N^{\mu}= n u^{\mu}$ is the particle flow vector, $u^{\mu}$ is the particle velocity, $\Theta= u^{\mu}_{;\mu}= 3 H$, stands for the fluid expansion, $\Gamma$ represents the particle creation rate, and, notationally, $\dot{\eta}= \eta_{;\mu} u^{\mu}$. The sign of $\Gamma$ indicates creation (for $\Gamma> 0$), or, annihilation (for $\Gamma< 0$) of particles, and, due to $\Pi$ which adds some dissipative effect to the cosmic fluid, non-equilibrium thermodynamics comes into picture.\\

Now, from the Gibb's equation using Clausius relation, we have \cite{Harko1}

\begin{equation}
Tds= d\left(\frac{\rho}{n}\right)+p d\left(\frac{1}{n}\right),\label{eqn5}
\end{equation}
where `$s$' represents entropy per particle, and $T$ is the fluid temperature. Using the conservation relations (\ref{conservation-eqn3}) and (\ref{eqn4}), the variation of entropy can be expressed as \cite{Zimdahl1, Chakraborty1}

\begin{equation}
n T \dot{s}= -\Pi \Theta- \Gamma\left(\rho+p\right).\label{eqn6}
\end{equation}
Further, for simplicity, if we assume the thermal process to be adiabatic (i.e., $\dot{s}= 0$),
then from Eq. (\ref{eqn6}) we have \cite{Zimdahl1, Chakraborty1}

\begin{equation}
\Pi= -\frac{\Gamma}{\Theta}\left(\rho+ p\right).\label{eqn7}
\end{equation}
Thus, the dissipative pressure is completely characterized by the particle creation rate for the above simple (isentropic) thermodynamical system. In other words, the cosmic substratum may be considered as a perfect fluid with barotropic equation of state, $p= (\gamma-1)\rho$, ($2/3 < \gamma \leq 2$), together with a dissipative phenomena which comes into picture through the particle creation mechanism. Furthermore, although the `entropy per particle' is constant, but still there is entropy generation due to particle creation, i.e., enlargement of the phase space through expansion of the universe. So, in some sense, the non-equilibrium configuration is not the conventional one due to the effective bulk pressure, rather, a state with equilibrium properties as well (but not the equilibrium era with $\Gamma= 0$). Now, eliminating $\rho$, $p$ and $\Pi$ from the Friedmann's equations (\ref{friedmann-eqn2}), the isentropic condition (\ref{eqn7}), and, using barotropic equation of state, $\gamma= 1+ p/\rho$, we obtain

\begin{equation}
\frac{\Gamma}{\Theta}= 1+\frac{2}{3\gamma}\frac{\dot{H}}{H^2}.\label{eqn8}
\end{equation}
The above equation shows that in case of adiabatic process, the particle creation rate is related to the evolution of the universe.

\section{Particle creation rate as a function of the Hubble parameter and evolution of the universe}

Introducing the deceleration parameter

\begin{equation}
q\equiv -\left(1+ \frac{\dot{H}}{H^2}\right),\label{eqn9}
\end{equation}
and, using Eq. (\ref{eqn8}), we have

\begin{equation}
q= -1+ \frac{3\gamma}{2}\left(1- \frac{\Gamma}{\Theta}\right).\label{eqn10}
\end{equation}
In the following subsections, we shall choose $\Gamma$ as different functions of the Hubble parameter to describe different stages of evolution of the universe and examine whether $q$ so obtained from Eq. (\ref{eqn10}) has any transition (from deceleration to acceleration or vice versa), or, not. Furthermore, it would be worthwhile to see the statefinder analysis for the particle creation rates in different stages of our universe. The statefinder parameters were introduced by Sahni el at. \cite{Sahni2003} as a geometrical concept to filter several observationally supported dark energy models from other phenomenological dark energy models existing in the literature. They introduced two new geometrical variables $r, s$ as follows \cite{Sahni2003}

\begin{eqnarray}
r&=& \frac{1}{a H^3} \dddot{a},~~~~~\mbox{and},~~s=\frac{r-1}{3 (q- 1/2)}.\label{statefinder1}
\end{eqnarray}

\begin{figure}
\begin{minipage}{0.4\textwidth}
\includegraphics[width= 1.0\linewidth]{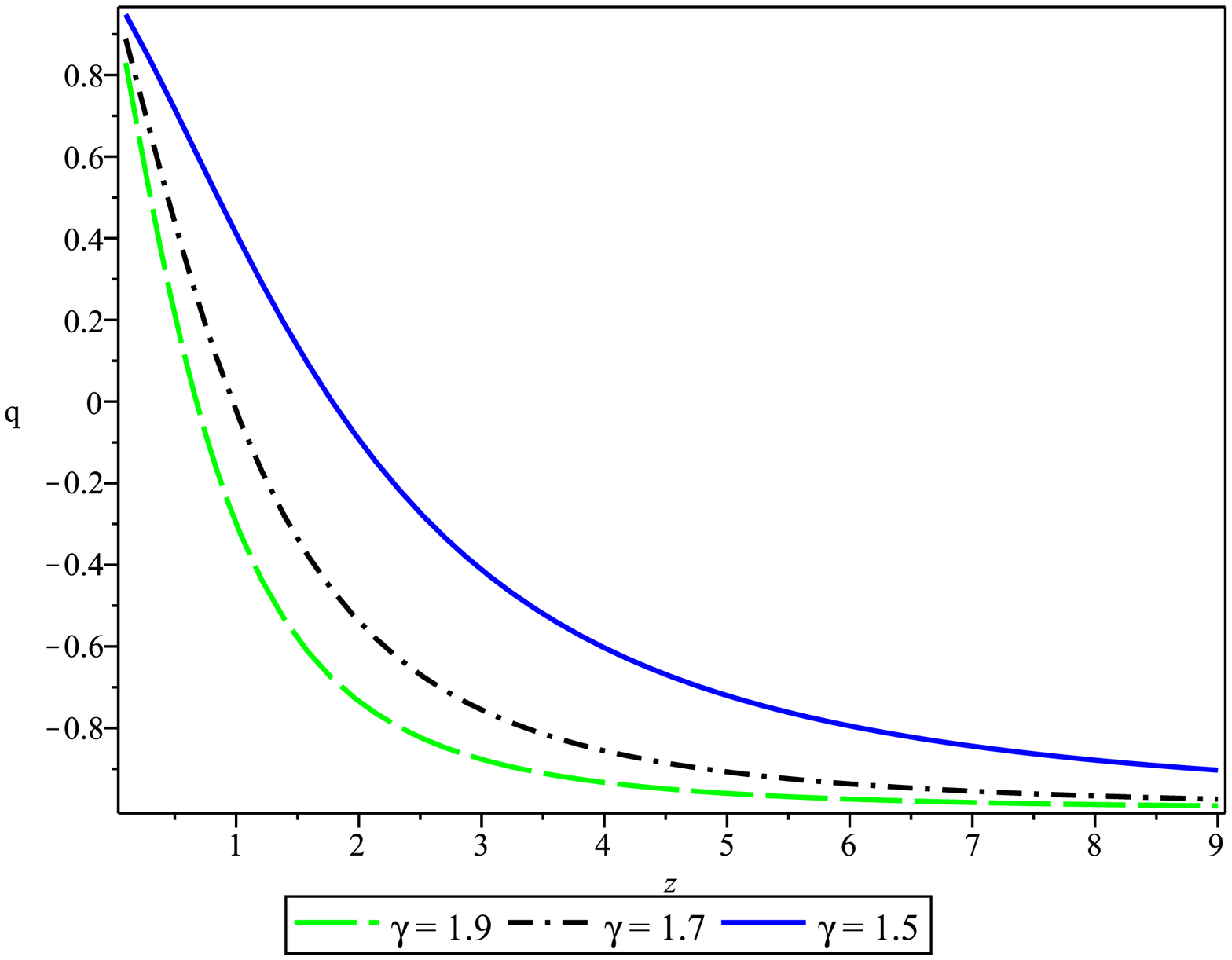}\\
Figure 1: The figure shows the transition from early inflationary phase $\rightarrow$ decelerating stage [see Eq. (\ref{eqn13})].
\end{minipage}
\end{figure}

\begin{figure}
\begin{minipage}{0.4\textwidth}
\includegraphics[width= 0.8\linewidth]{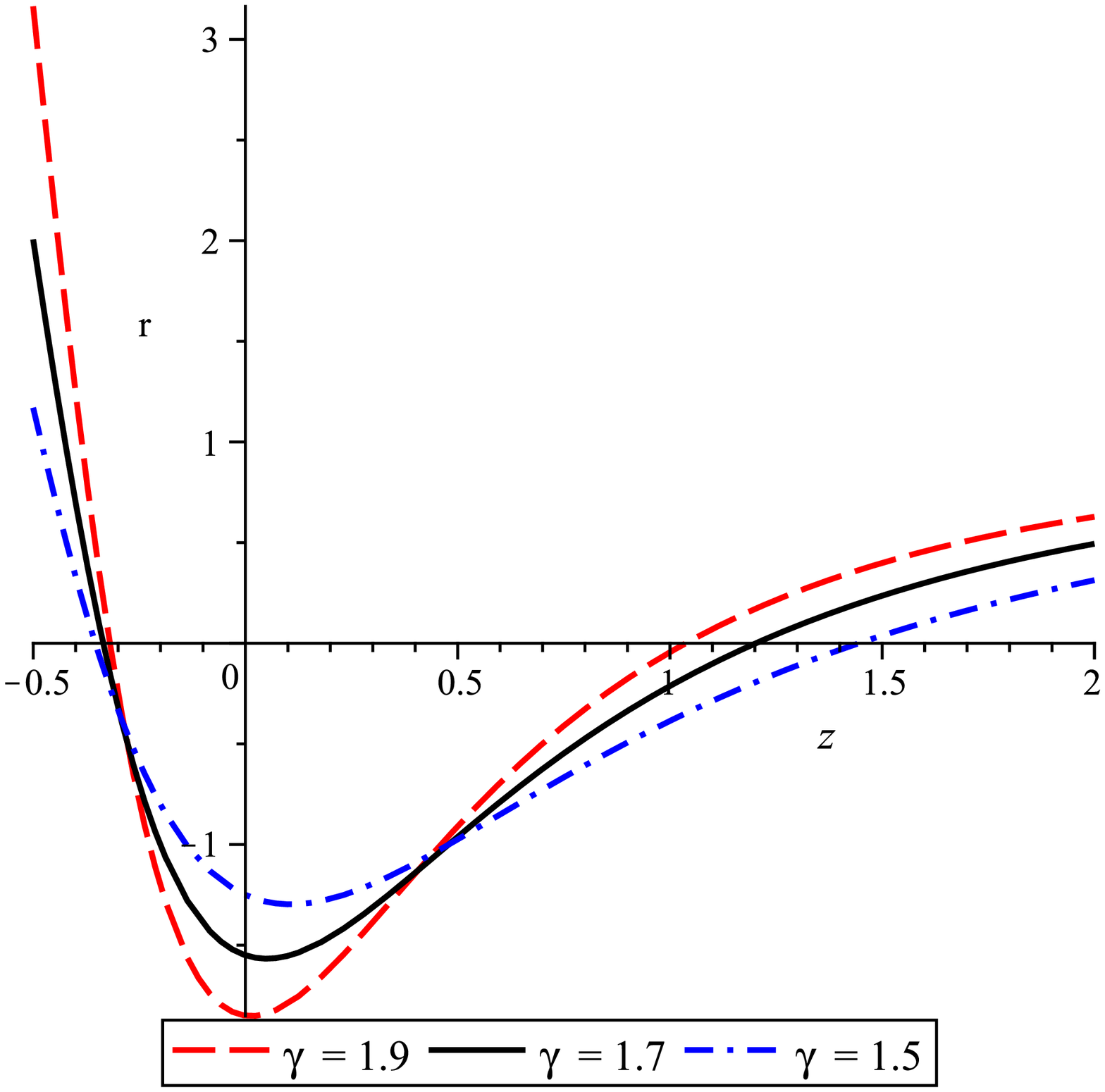}\\
Figure 1.1: The figures show the variation of $r$ against $z$ in early stage of the universe for three different choices of $\gamma$.
\end{minipage}
\begin{minipage}{0.4\textwidth}
\includegraphics[width= 0.8\linewidth]{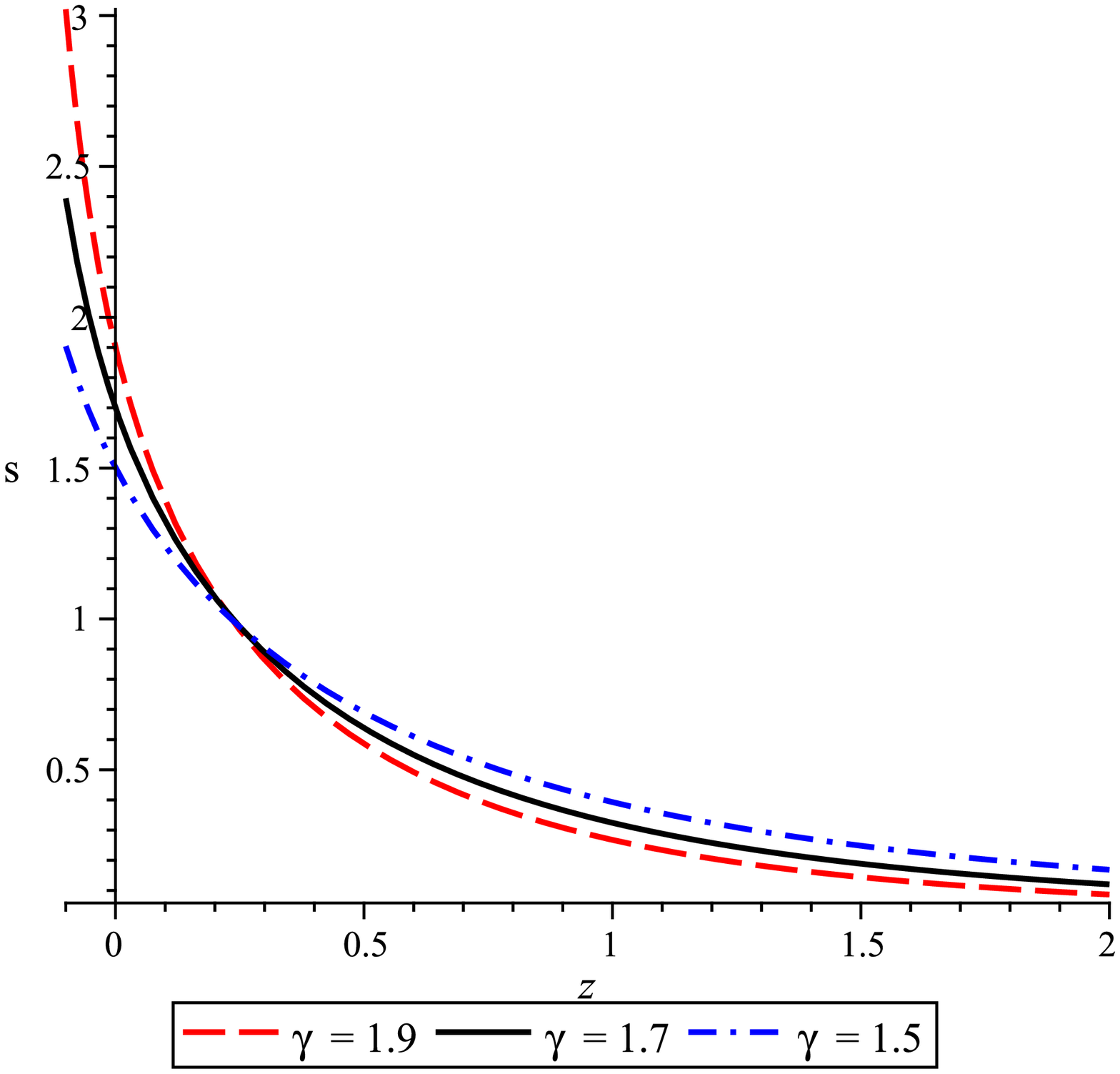}\\
Figure 1.2: The second statefinder parameter $s$ is shown over $z$ for three different choices of $\gamma$.
\end{minipage}
\end{figure}

\subsection{Early epochs}
\label{Early-phase}

In the very early universe (starting from a regular vacuum) most of the particle creation effectively takes place and from thermodynamic point of view we have \cite{Lima3}\\

(i) At the beginning of the expansion, there should be maximal entropy production rate (i.e., maximal particle creation rate) so that universe evolves from non-equilibrium thermodynamical state to equilibrium era with the expansion of the universe.\\

(ii) A regular (true) vacuum for radiation initially, i.e., $\rho \longrightarrow 0$, as $a \longrightarrow 0$.\\

(iii) $\Gamma > H$ in the very early universe so that the created radiation behaves as thermalized heat bath and subsequently, the creation rate should fall slower than expansion rate and particle creation becomes dynamically insignificant.\\

Now, according to Gunzig et al. \cite{Gunzig1}, the simplest choice satisfying the above requirements is that particle creation rate is proportional to the energy density, i.e., $\Gamma= \Gamma_0 H^2$ (where $\Gamma_0$ is a proportionality constant). For this choice of $\Gamma$, $H$ can be solved from (\ref{eqn8}) as

\begin{equation}
H= \frac{H_e}{\beta+(1-\beta)(a/a_e)^{\frac{3\gamma}{2}}},\label{eqn11}
\end{equation}
where $\beta$ is related to $\Gamma_0$ as $\Gamma_0= 3 \beta/H_e$. $H_e$, and $a_e$ are chosen to be the values of the Hubble parameter and the scale factor respectively at some instant. We note that as $a\longrightarrow 0, H\longrightarrow \beta^{-1}H_e=$ constant, indicating an exponential expansion ($\ddot{a}> 0$) in the inflationary era, while for $a\gg a_e$, $H\propto a^{-3\gamma/2}$, indicates the standard FLRW cosmology ($\ddot{a}< 0$). So, if we identify ``$a_e$" at some intermediate value of `$a$', where, $\ddot{a}= 0$, i.e., a transition from de Sitter accelerating phase to the standard decelerating radiation phase, then we have $\dot{H_e}= -H_e ^2$, and, Eq. (\ref{eqn8}) gives, $\beta= 1-2/3 \gamma$.\\


Now, using Eq. (\ref{eqn10}), we obtain

\begin{equation}
q(z)= -1+\frac{3\gamma}{2}\left [1-\frac{\beta}{\beta+(1-\beta)(1+z)^{-3\gamma /2}}\right],\label{eqn13}
\end{equation}
where the redshift parameter is defined as: $a_e/a= 1+ z$.\\

Figure 1 shows the transition of the deceleration parameter $q(z)$ from early inflationary era to the decelerated radiation era. Figures 1.1 and 1.2 display
the graphical behavior of the statefinder parameters showing that in early stage of the universe, $r$ changes its sign from $+$ve ($\equiv$ inflationary era) to $-$ve ($\equiv$ decelerating phase), while $s$ stays positive throughout the transition.

\subsection{Intermediate decelerating phase}

Here the simple natural choice is $\Gamma \propto H$, i.e., $\Gamma= \Gamma_1 H$ (where $\Gamma_1$ is the proportionality constant). It should be noted that this choice of $\Gamma$ does not satisfy the third thermodynamical requirement (mentioned above) at the early universe. Also the solution will not satisfy the above condition (ii) of \ref{Early-phase}. In this case $q$ does not depend on the expansion rate, it only depends on $\gamma$. For radiation (i.e., $\gamma= 4/3$) era, $q= 1-\frac{2\Gamma_1}{3}$, while for matter dominated era (i.e., $\gamma= 1$) $q= \frac{1}{2}-\frac{\Gamma_1}{2}$. So, if $\Gamma_1< 1$, then we have deceleration in both the epochs as in standard cosmology, while there will be acceleration, if $\Gamma_1 > (3-\frac{2}{\gamma})$.\\

The solution for the Hubble parameter and the scale factor are given by

\begin{eqnarray}
H^{-1}&=& \frac{3\gamma}{2}\left(1-\frac{\Gamma_1}{3}\right)t,~~~\mbox{and},~~~a=a_0 t^l,\label{eqn13.1}
\end{eqnarray}
where $l= 2/3\gamma (1-\frac{\Gamma_1}{3})$, which represents the usual power law expansion of the universe in standard cosmology with particle production rate decreases as $t^{-1}$.\\

The statefinder parameters $r$ and $s$ in this case are constant with the values

\begin{eqnarray}
r&=& \left(1-\frac{1}{l}\right)\left(1- \frac{2}{l}\right),~~~~s=\frac{2}{3}\left(1-\frac{1}{l}\right)\left(1- \frac{2}{l}\right) \left(\frac{2}{l}-3\right)^{-1},\label{statefinder2}
\end{eqnarray}
which indicates that, for $l \longrightarrow \infty$, $r$ and $s$ changes its sign. $r$ becomes $+$ve, while $s$ becomes $-$ve.

\begin{figure}
\begin{minipage}{0.4\textwidth}
\includegraphics[width= 1.0\linewidth]{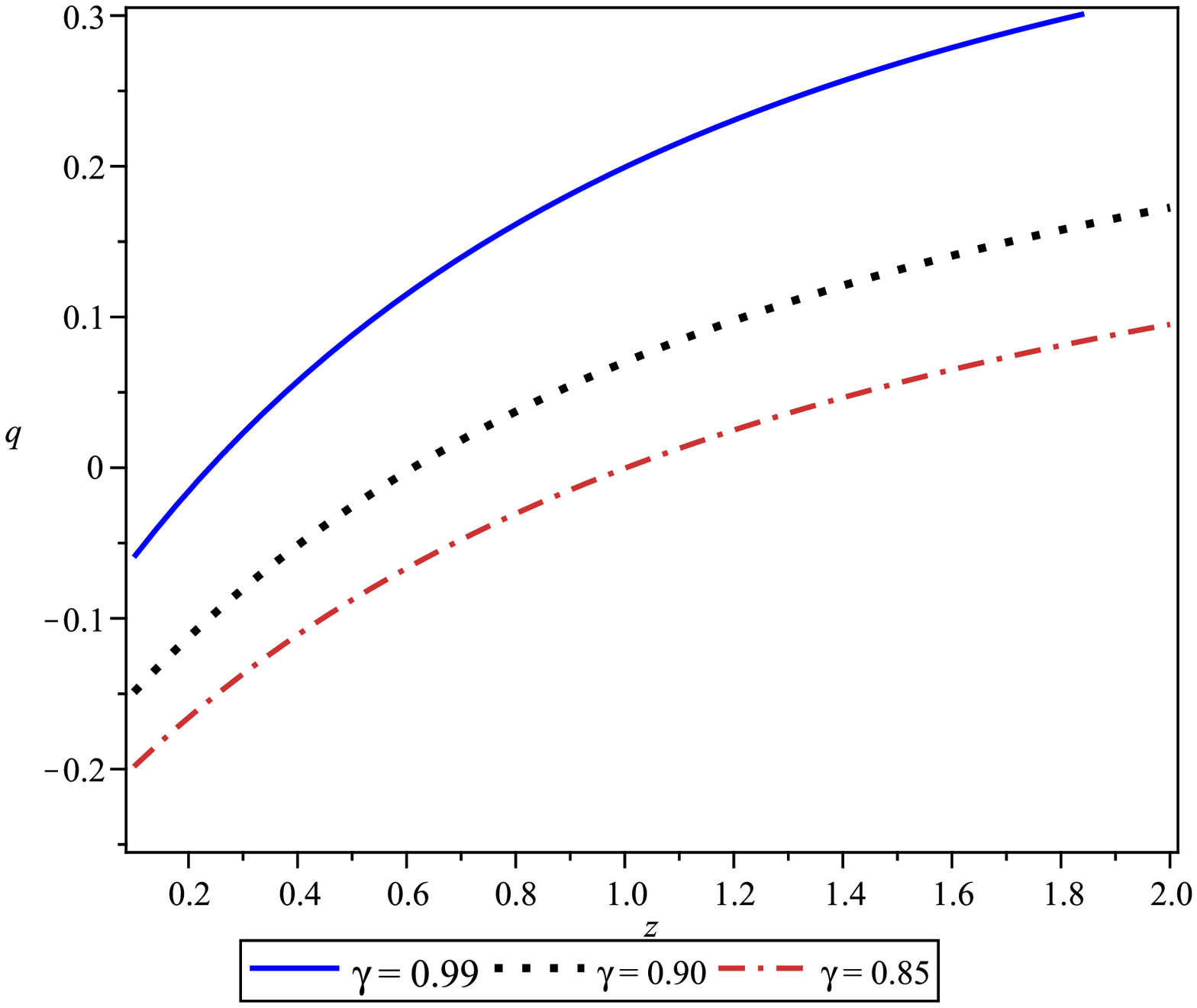}\\
Figure 2: It describes the cosmic evolution from decelerating phase $\rightarrow$ recent accelerating stage [see Eq. (15)].
\end{minipage}
\end{figure}

\begin{figure}
\begin{minipage}{0.4\textwidth}
\includegraphics[width= 0.8\linewidth]{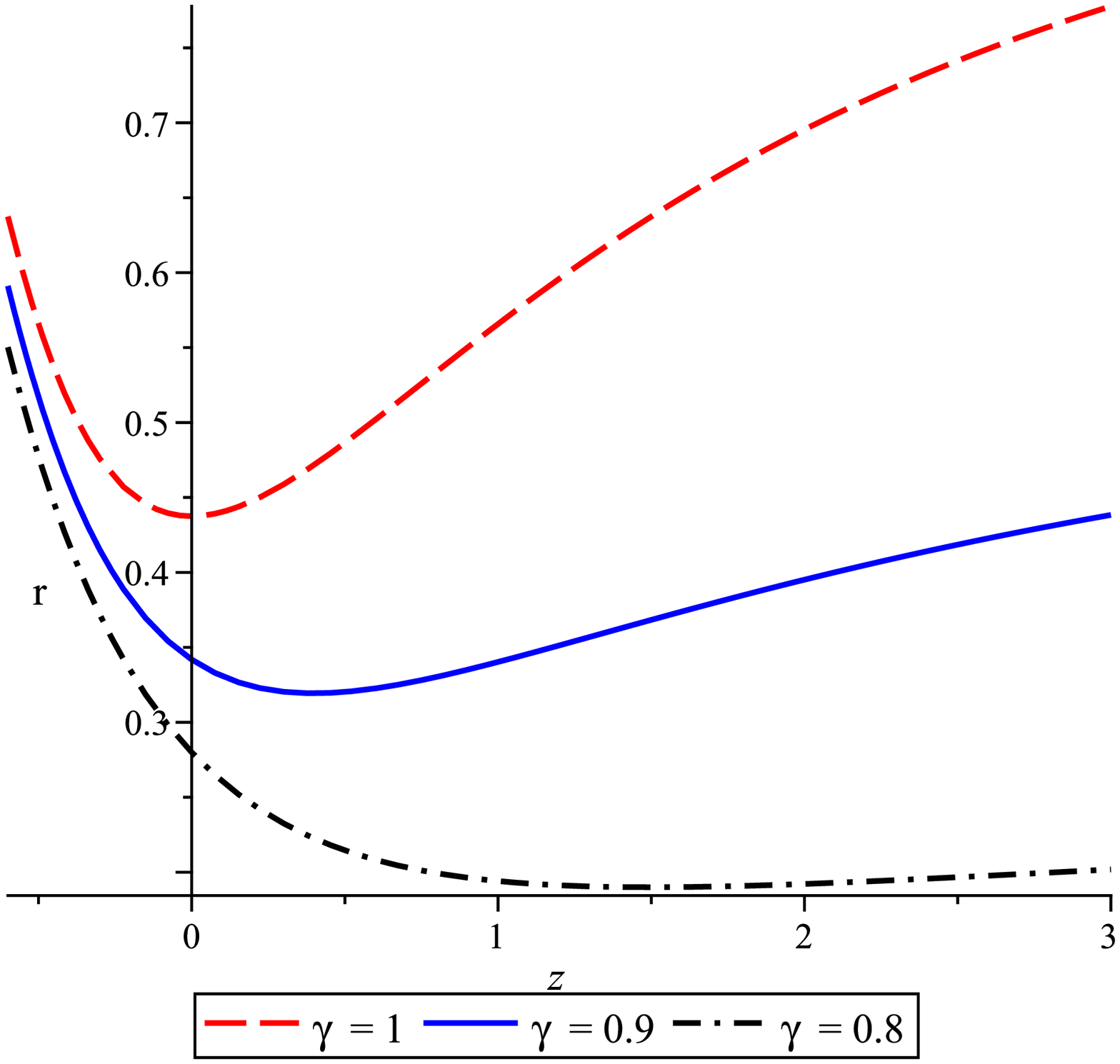}\\
Figure 2.1: In late-time, this is the variation of $r$ over $z$.
\end{minipage}
\begin{minipage}{0.4\textwidth}
\includegraphics[width= 0.8\linewidth]{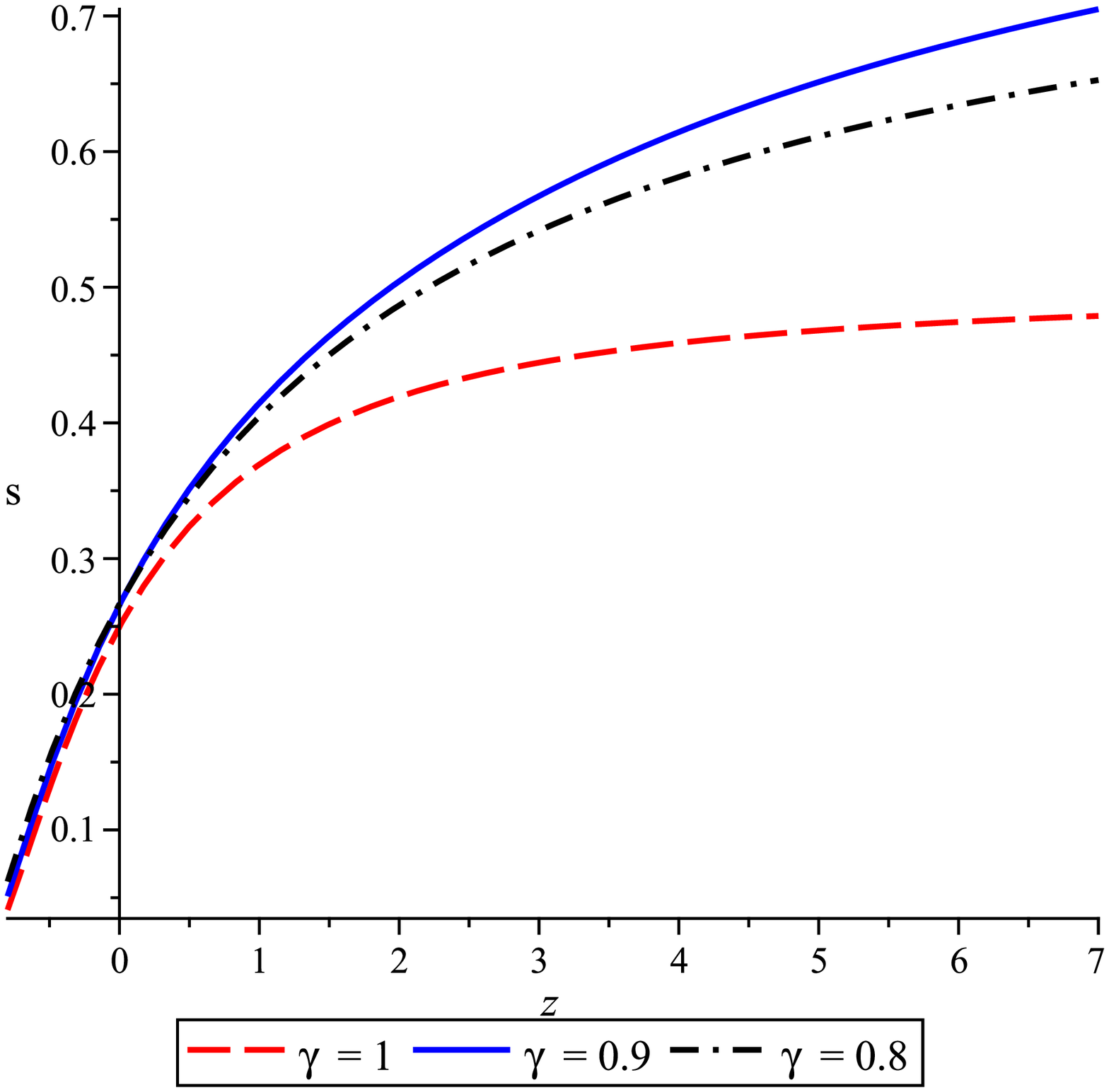}\\
Figure 2.2: The figure shows the behavior of $s$ for three different values for $\gamma$.
\end{minipage}
\end{figure}

\subsection{Late time evolution: Accelerated expansion}

In this case the thermodynamical requirements of \ref{Early-phase} are modified as \cite{Gunzig1}\\

(i) There should be minimum entropy production rate at the beginning of the late time accelerated expansion and the universe again becomes non-equilibrium thermodynamically.\\

(ii) The late time false vacuum should have $\rho$ $\longrightarrow$0, as $a \longrightarrow \infty$.\\

(iii) The creation rate should be faster than the expansion rate.\\

We shall show that another simple choice of $\Gamma$, namely, $\Gamma \propto 1/H$, i.e., $\Gamma=\Gamma_3/H$, where, $\Gamma_3$ is a proportionality constant, will satisfy these requirements. For this choice of $\Gamma$, the Hubble parameter is related to the scale factor as

\begin{equation}
H^2= \frac{\Gamma_3}{3}+(a/a_f)^{-3\gamma},\label{eqn14}
\end{equation}
where, $a_f$ is some intermediate value of `$a$', such that

\begin{eqnarray}
H&\sim& a^{-3\gamma/2},~~~~~~\mbox{for,}~~~~~~~a\ll a_f,\label{eqn14.1}\\
H&\sim& \frac{\Gamma_3}{3},~~~~~~~~~~\mbox{for,}~~~~~~~a\gg a_f.\label{eqn14.2}
\end{eqnarray}
So, we have a transition from the standard cosmological era ($\ddot{a}< 0$) to late time acceleration ($\ddot{a}> 0$), and, $a_f$'s can be identified as the value of the scale factor at the instant of transition ($\ddot{a}= 0$). The deceleration parameter now has the expression

\begin{equation}
q= -1+ \frac{3\gamma}{2}\left[\frac{1}{1+(\frac{\Gamma_3}{3})(1+z)^{-\frac{3\gamma}{2}}}\right],\label{eqn15}
\end{equation}
where the redshift parameter is defined as: $a_f/a= 1+ z$. Figure 2 displays the transition
of the universe from matter dominated era to the present late-time acceleration.
Further, we have presented the statefinder analysis in Figures 2.1 and 2.2 for
$r$ and $s$ parameters respectively.

\section{Field Theoretic Analysis and Particle Creation}

This section deals with particle creation from vacuum using quantum field theory \cite{Graef1}. In particular, the quantum effect of particle creation is considered in the context of thermodynamics of open systems and is interpreted as an additional negative pressure.\\

The energy-momentum tensor corresponding to the quantum vacuum energy is

\begin{eqnarray}
T^Q_{\mu\nu}\equiv <T^Q_{\mu\nu}>= \Lambda(t)g_{\mu\nu}.\label{field-theory}
\end{eqnarray}
So, the Friedmann's equations, and, the corresponding energy-conservation equation of a perfect fluid are now modified as

\begin{eqnarray}
3 H^2&=& 8 \pi G (\rho+ \Lambda),\label{eqn16}\\
2 \dot{H}&=& -8 \pi G \left(p+ \rho\right),\label{eqn17}
\end{eqnarray}
and,

\begin{equation}
\dot{\rho}+3H(p+\rho)= -\dot{\Lambda},\label{eqn18}
\end{equation}
which shows a energy transfer from the decaying vacuum to matter. This modified energy conservation equation can be considered as the energy balance equation for an imperfect fluid with bulk viscous pressure

\begin{eqnarray}
\Pi= \frac{\dot{\Lambda}}{3H}.\label{Pi}
\end{eqnarray}
Now, if the perfect fluid is considered as a scalar field with potential $V(\phi)$, i.e.,

\begin{eqnarray}
\rho_\phi= \frac{1}{2}\dot{\phi}^2+V(\phi),\label{rho-field}\\
p_\phi= \frac{1}{2}\dot{\phi}^2-V(\phi),\label{p-field}
\end{eqnarray}
then the Einstein's field Eqns. (\ref{eqn16}) and (\ref{eqn17}) become respectively (taking $8 \pi G= 1$)

\begin{equation}
3H^2= \frac{1}{2}\dot{\phi}^2+ V(\phi)+\Lambda(t),~~~~~~~~~~~2\dot{H}= -\dot{\phi}^2,\label{eqn19}
\end{equation}
and, the evolution Eq. (\ref{eqn18}) of the scalar field is given by

\begin{equation}
\dot{\phi} \ddot{\phi}+ \dot{\phi}\frac{dV}{d\phi}+3 H \left(\dot{\phi}^2+\frac{\dot{\Lambda}}{3H}\right)= 0.\label{eqn20}
\end{equation}
So, we have

\begin{equation}
\phi= \int \sqrt{\frac{-2H^\prime}{aH}},~~~~~~~\mbox{and},~~~~~~~~~ V= -\Lambda+3 H^2 \left[1+\frac{aH^\prime}{3H}\right],\label{eqn21}
\end{equation}
where `$^\prime$' stands for the differentiation with respect to the scale factor `$a$'. Hence, for adiabatic process the particle creation rate can be written as

\begin{equation}
\Gamma= \frac{H}{2(1+q)}\left [(4-r)+\frac{1}{H^3}\frac{dV}{d\phi}\sqrt{-2\dot{H}}\right],\label{eqn22}
\end{equation}
where $r= \frac{\dddot{a}}{aH^3}$ is the state finder parameter \cite{Sahni2003}, and $q= -(1+\frac{\dot{H}}{H^2})$ is the usual deceleration parameter.\\

It may be noted that, if we have only quantum energy and there is no other matter, then the energy conservation equation (\ref{eqn18}) demands `$\Lambda$' should be a constant. Also, from Eq. (\ref{eqn8}) we have constant particle creation rate $\sqrt{3 \Lambda}$.

\section{Two non-interacting fluids as cosmic substratum and particle creations}

\begin{figure}
\begin{minipage}{0.4\textwidth}
\includegraphics[width= 1.0\linewidth]{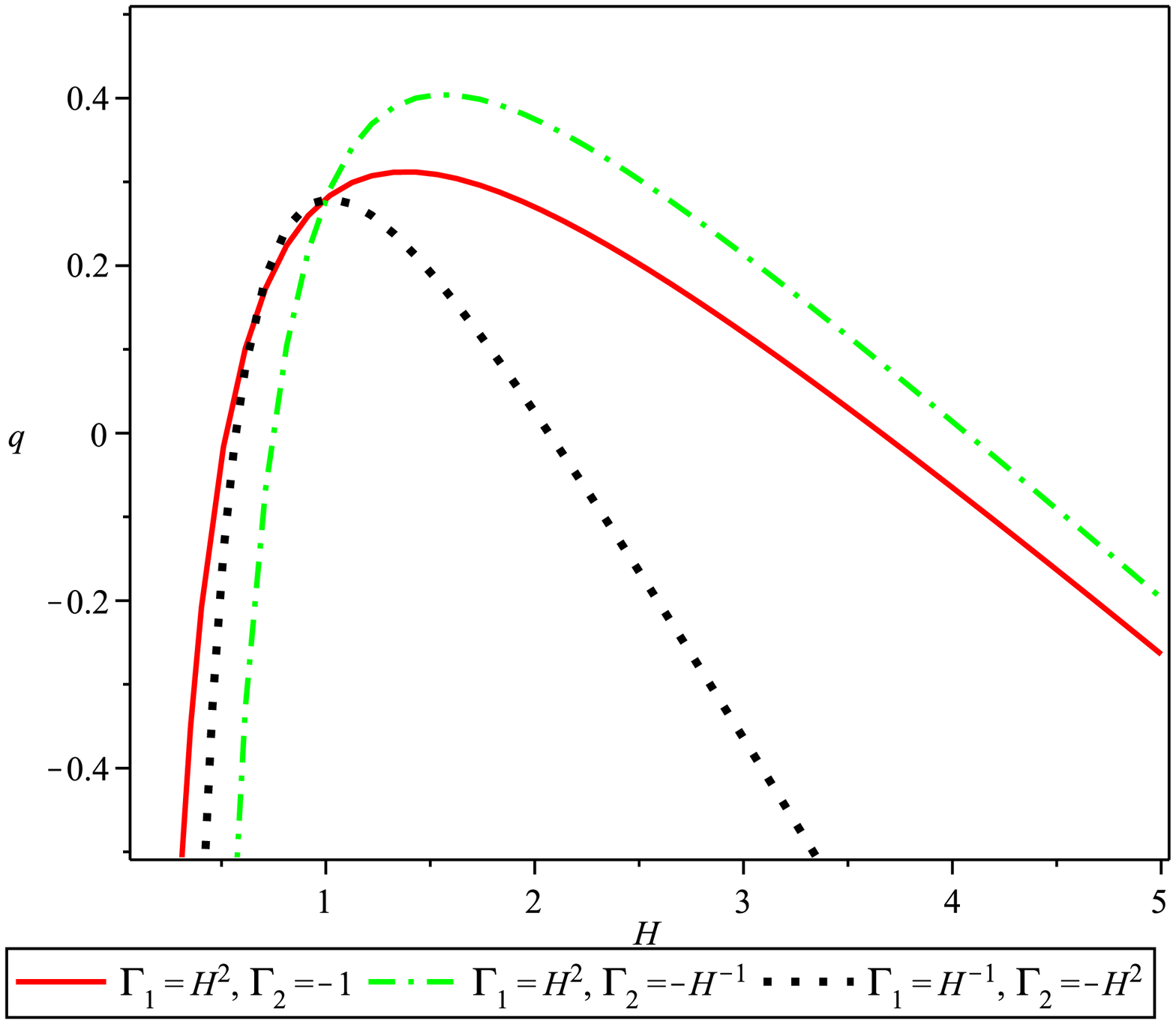}
Figure 3: It is a comparative study of the deceleration parameter ($q$) with the Hubble parameter ($H$) for different particle creation rates ($\Gamma$) for the set of values $\Omega_1= 0.4$, $\Omega_2= 0.6$, $\omega_1= 0.1$, and  $\omega_2= 0.4$.
\end{minipage}
\begin{minipage}{0.4\textwidth}
\includegraphics[width= 1.0\linewidth]{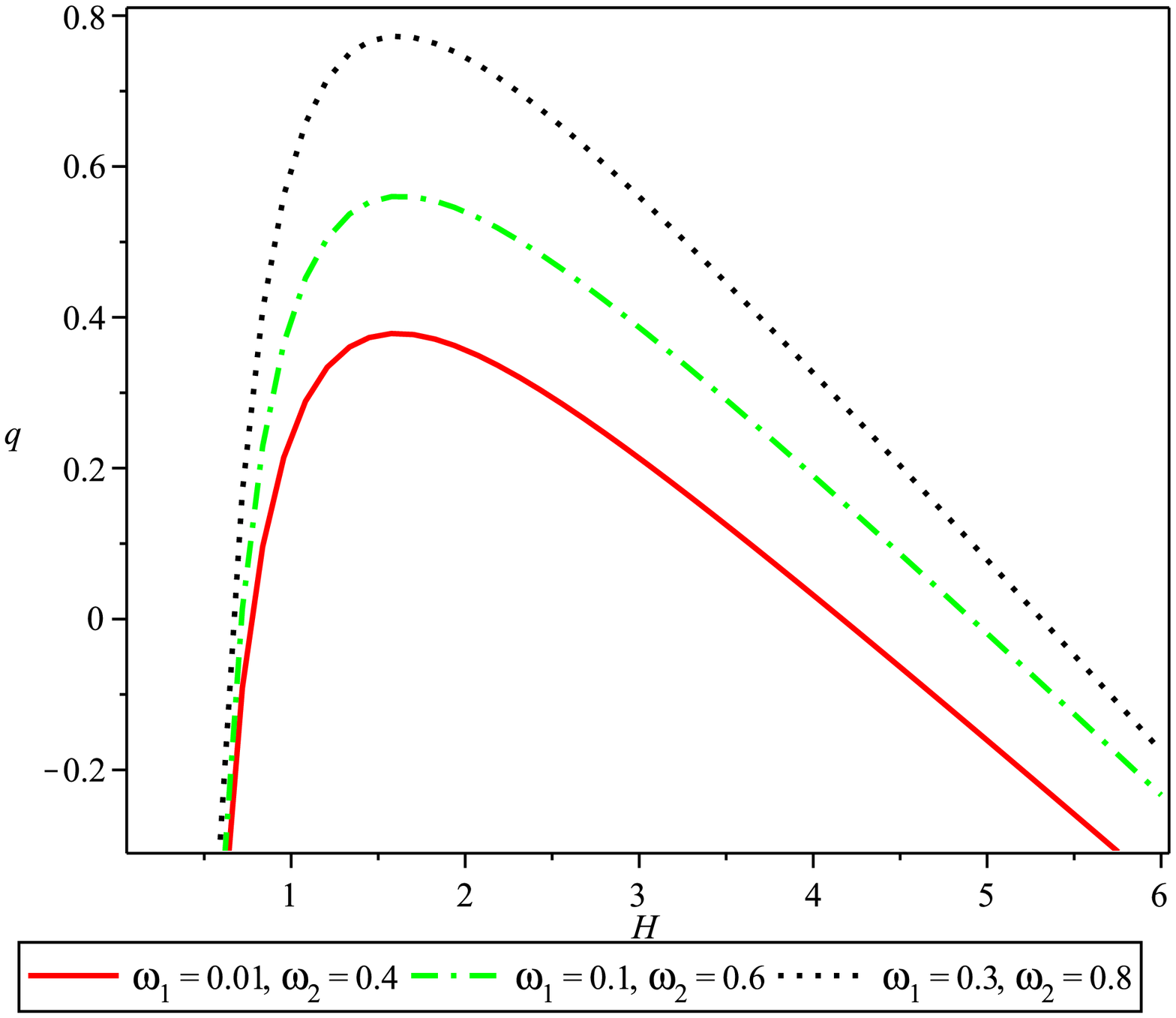}
Figure 4: The figure shows the variation of the deceleration parameter ($q$) with the Hubble parameter ($H$) in different equation of states for the following unequal particle creation rates: $\Gamma_1= H^2$, $\Gamma_2= -\frac{1}{H}$, for $\Omega_1= 0.4, \Omega_2= 0.6$.
\end{minipage}
\end{figure}
\begin{figure}
\begin{minipage}{0.4\textwidth}
\includegraphics[width= 1.1\linewidth]{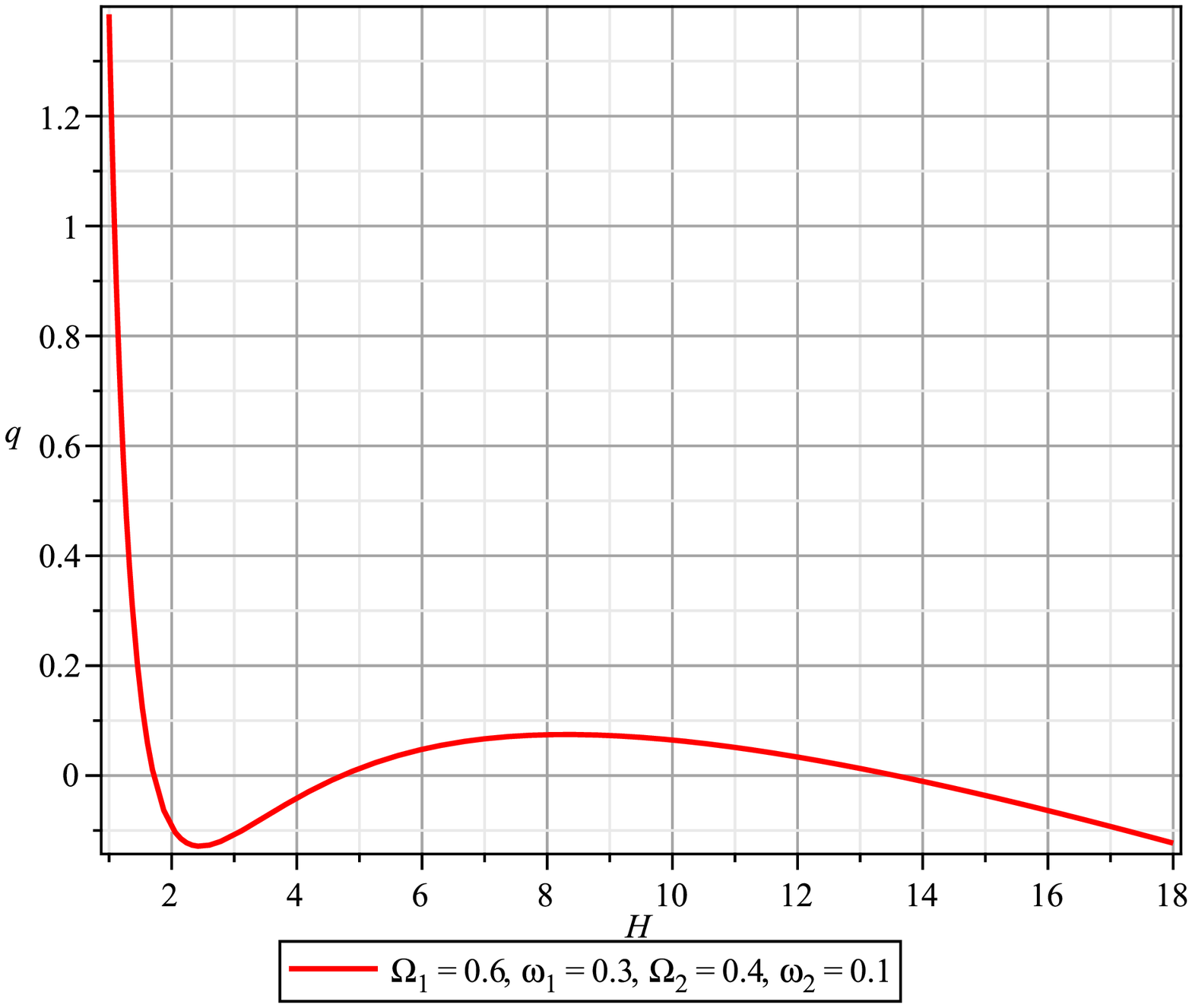}
Figure 5: The variation of $q$ against $H$ for unequal PCR has been shown in this way: inflation $\rightarrow$ deceleration $\rightarrow$ late time acceleration $\rightarrow$ future deceleration. Here, $\Gamma_1 \approx 0.12 H^2$ and $\Gamma_2 \approx 20.71/H- 18$.
\end{minipage}
\end{figure}

In this section, we suppose that the present open thermodynamical system contains two non-interacting dark fluids which have different particle creation rates. Let, ($\rho_1, p_1$) and ($\rho_2, p_2$) are the energy density and thermodynamic pressure of the fluids respectively. Suppose ($n_1, n_2$) denote the number density of the two fluids having balance equations \cite{Harko1}

\begin{equation}
\dot{n_1}+3H n_1= \Gamma_1 n_1,~~~~~~\mbox{and},~~~~~~~~~\dot{n_2}+ 3H n_2= -\Gamma_2 n_2,\label{eqn23}
\end{equation}
where $\Gamma_1> 0$ and $\Gamma_2> 0$. The above equations imply that, there is creation of particles of fluid-1, while particles of fluid-2 decay. Now, combining equations in (\ref{eqn23}), the total number of particles, $n= n_1+ n_2$, will have the balance equation

\begin{equation}
\dot{n}+3H n= \left(\frac{\Gamma_1 n_1-\Gamma_2 n_2}{n}\right)n= \Gamma n.\label{eqn24}
\end{equation}
So, the total number of particles will remain conserve, if $\Gamma= 0$, i.e., $\Gamma_1 n_1= \Gamma_2 n_2$. Again from the isentropic condition in Eq. (\ref{eqn7}), the dissipative (bulk) pressure of the matter components are given by

\begin{equation}
\Pi_1= -\frac{\Gamma_1}{3H}(\rho_1+ p_1),~~~~~~~~\mbox{and},~~~~~~~~\Pi_2= -\frac{\Gamma_2}{3H}(\rho_2+ p_2).\label{eqn25}
\end{equation}
As a consequence, the energy conservation relations are
\begin{equation}
\dot{\rho_1}+3H(\rho_1+ p_1)= \Gamma_1(\rho_1+p_1),~~~~\mbox{and},~~~~~
\dot{\rho_1}+3H(\rho_1+ p_1)= -\Gamma_2(\rho_2+p_2),\label{eqn26}
\end{equation}
which imply an exchange of energy between the two fluids. In this connection, it should be mentioned that Barrow and Clifton \cite{Barrow2} found cosmological solutions with energy exchange. Now, if $\omega_1= \frac{p_1}{\rho_1}$ and $\omega_2= \frac{p_2}{\rho_2}$ are the equations of state of the two fluid components respectively, then from the above two conservation relations the effective equation of state parameters are

\begin{equation}
w^{eff} _1= \omega_1-\frac{\Gamma_1}{3H}(1+\omega_1),~~~~~\mbox{and},~~~~~w^{eff} _2= \omega_2+\frac{\Gamma_2}{3H}(1+\omega_2).\label{eqn27}
\end{equation}
Thus, from the Einstein's equations we have

\begin{equation}
3 H^2= \rho_1+ \rho_2,~~~~~\mbox{and},~~~~~2\dot{H}= -\left[(\rho_1+ p_1+ \Pi_1)+ (\rho_2+ p_2+ \Pi_2)\right].\label{eqn28}
\end{equation}
Then the deceleration parameter can be written as \cite{Pan1}

\begin{figure}
\begin{minipage}{0.4\textwidth}
\includegraphics[width= 1.0\linewidth]{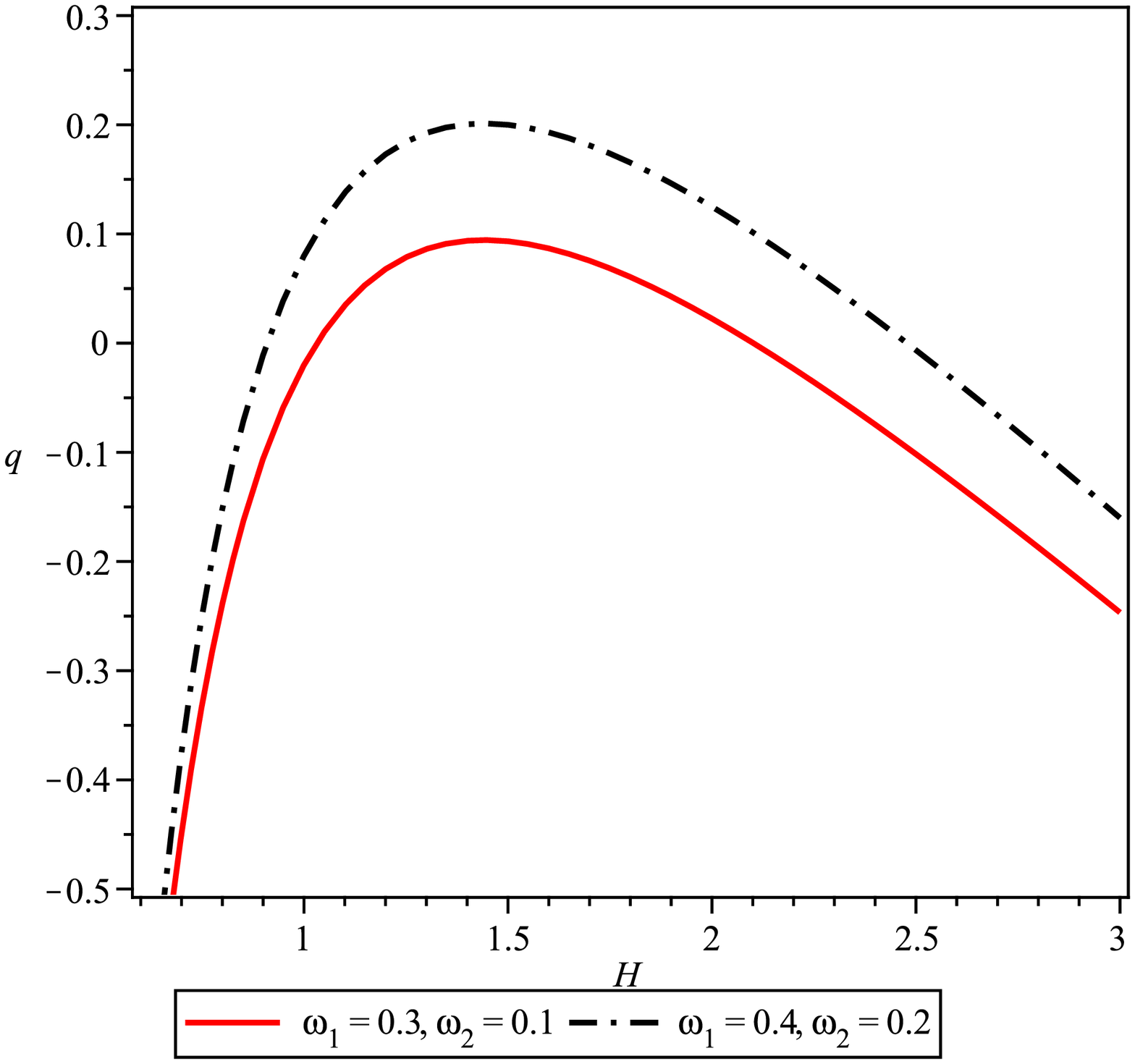}\\
Figure 6: The evolution from inflation $\rightarrow$ deceleration $\rightarrow$ late time acceleration for a single PCR, $\Gamma= 2 H^2+ 3/H$ has been shown. Here, $\Omega_1= 0.6$ and $\Omega_2= 0.4$.
\end{minipage}
\begin{minipage}{0.4\textwidth}
\includegraphics[width= 1.0\linewidth]{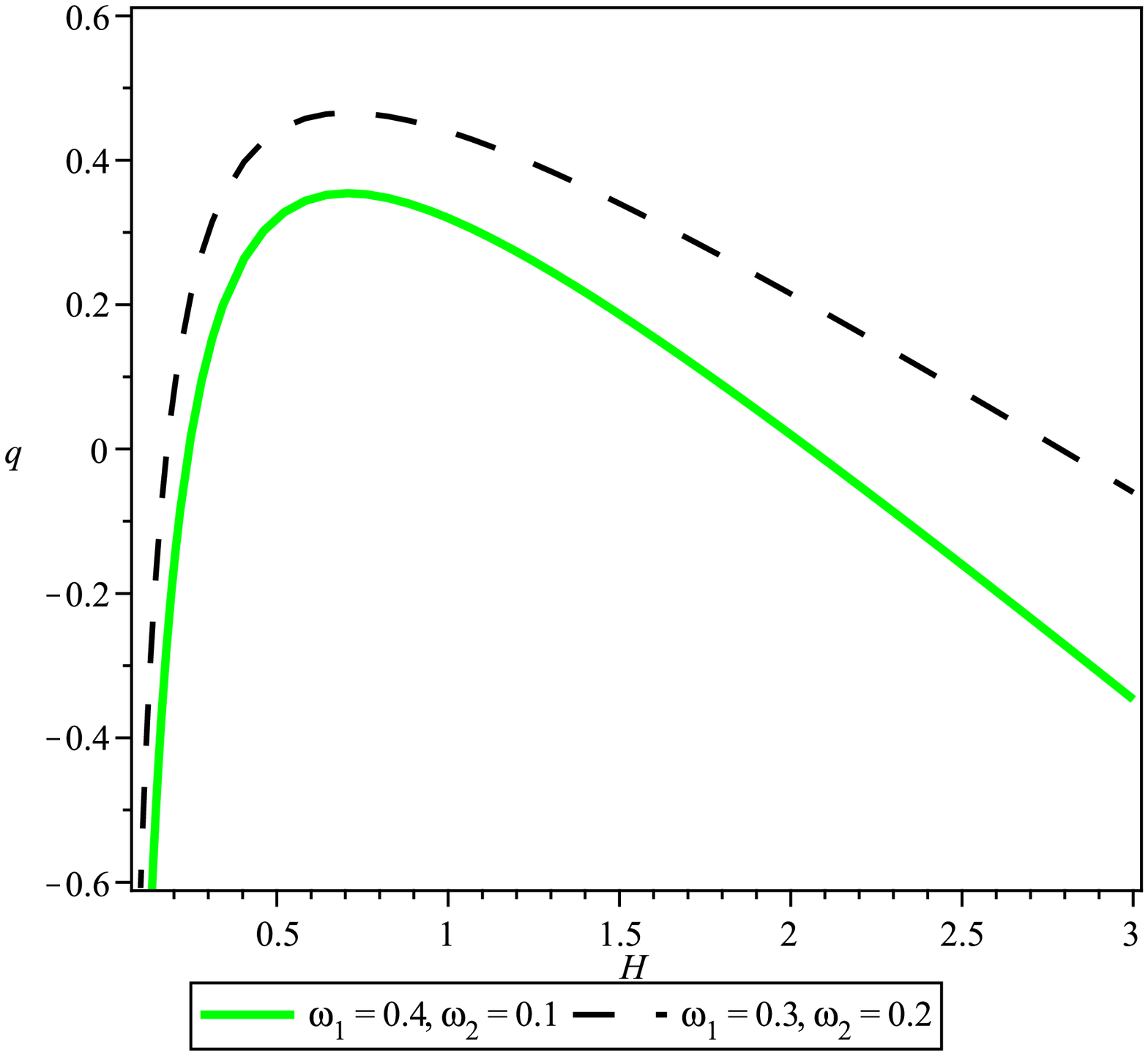}\\
Figure 7: It describes our universe as: inflation $\rightarrow$ deceleration $\rightarrow$ late-time acceleration for the particle creation rate $\Gamma= 2H^2+ 1$, with $\Omega_1= 0.6$, $\Omega_2= 0.4$.
\end{minipage}
\end{figure}
\begin{figure}
\begin{minipage}{0.4\textwidth}
\includegraphics[width= 1.1\linewidth]{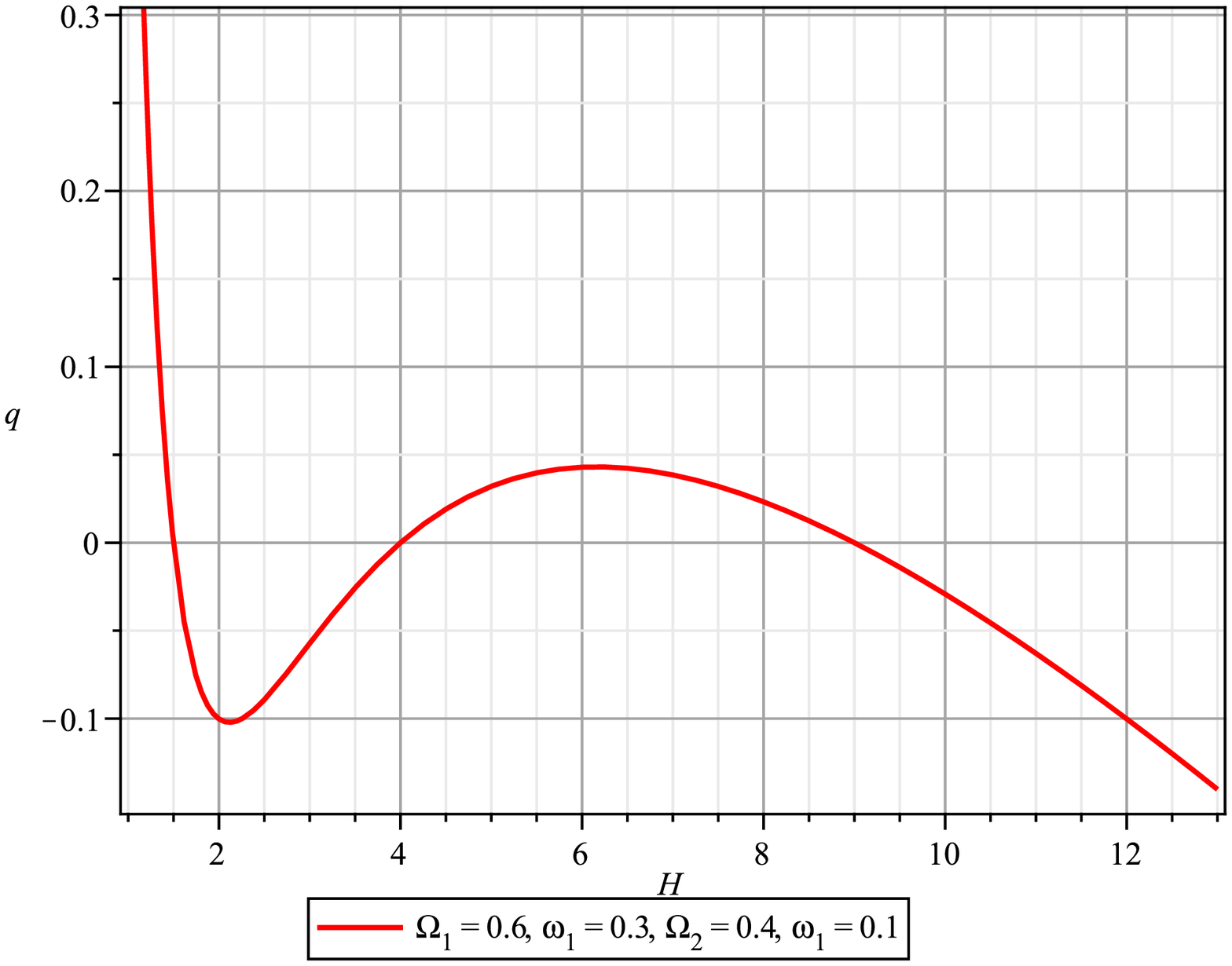}\\
Figure 8: The figure shows that for equal PCR, the complete scenario from inflation to present late time acceleration can be described, and not only that, it predicts a possible future deceleration of the universe. Here, $\Gamma \approx 0.34 H^2-18.18/H+ 18.69$.
\end{minipage}
\end{figure}

\begin{equation}
q= \frac{1}{2}+\frac{3}{2} \left[-\frac{\Gamma_1}{3H}\Omega_1(1+\omega_1)+\frac{\Gamma_2}{3H}\Omega_2(1+\omega_2)+(\Omega_1\omega_1+\Omega_2\omega_2)\right].\label{eqn29}
\end{equation}
In particular, if $\Gamma_1= \Gamma_2= \Gamma$ (say), then the above form of the deceleration parameter ($q$) reads

\begin{equation}
q= \frac{1}{2}+\frac{3}{2} \left[(\Omega_1 \omega_1+\Omega_2 \omega_2)+\frac{\Gamma}{3H}((\Omega_2 \omega_2-\Omega_1 \omega_1)+(\Omega_2-\Omega_1))\right].\label{eqn30}
\end{equation}
Figures 3--8 describe how the deceleration parameter behaves with the Hubble parameter for different choices of the particle creation rate (equal or unequal), and it also matches with the present day observations. It should be mentioned that the choice for $\Gamma_1$, and $\Gamma_2$ in figure 5 are not fixed, rather the cosmic evolution can be obtained for any $\Gamma_1= A H^2$, and $\Gamma_2= \frac{B}{H}+ C$ (where, $A, B, C$ are constants). Similarly, any PCR of the form (for figure 8), $\Gamma= D H^2+ \frac{E}{H}+ F$ (where, $D, E, F$ are constants) we can have the same evolution of the universe.

\section{Discussions and future prospects}
The present work deals with non-equilibrium thermodynamics based on particle creation formalism. In the context of universal thermodynamics, flat FLRW model of the universe is considered as the open thermodynamical system. Although, cosmic fluid is chosen in the form of a perfect fluid, but dissipative effect in the form of bulk viscous pressure arises due to the particle production mechanism. For simplicity of calculations, we are restricted to the adiabatic process where the dissipative pressure is linearly related to the particle production rate ($\Gamma$). From thermodynamic point of view, $\Gamma$ is chosen as a function of the Hubble parameter ($H$), and, the deceleration parameter is shown to be a function of the redshift parameter. In particular, by proper choices of $\Gamma$, cosmological solutions are evaluated, and, the deceleration parameter is presented graphically in figures 1 and 2. The graphs show a transition from early inflationary stage to the radiation dominated era (Figure 1), and, also the transition from matter dominated era to the present late time acceleration (Figure 2). Further, we have presented the statefinder analysis in early (Figures 1.1 and 1.2), intermediate ($r$ and $s$ are constant in this stage) and the late phases (Figures 2.1 and 2.2) respectively. Then a field theoretic analysis has been shown for the particle creation mechanism in section IV. Finally, in section V, a combination of non-interacting two perfect fluids having different particle creation rate is considered as a cosmic substratum, and the deceleration parameter is evaluated. The behavior of the deceleration parameter is examined graphically in Figures 3--8. Figures 3, 4, 6 and 7 show two transitions of $q$---one in the early epoch from acceleration to deceleration, and, the other one corresponds to the transition in the recent past from deceleration to present accelerating stage. The Figures 3 and 4 correspond to two different choices of unequal particle creation parameters, while for two distinct equal particle creation rate, the variation of $q$ are presented in Figures 6 and 7. There are three distinct transitions of $q$ for unequal and equal particle creation parameters in Figures 5 and 8 respectively. Both the figures show that there is a chance of our universe to decelerate again in future from the present accelerating stage. Thus, theoretically, considering non-interacting two fluid system as cosmic substratum, it is possible to have again a decelerating phase of the universe in future. Therefore, we may conclude that the present observed accelerating phase is due to non-equilibrium thermodynamics having particle creation processes, or, in other words, in addition to the presently known two possibilities (namely,  modification of Einstein gravity, or, introduction of some unknown exotic fluid, i.e., DE) for explaining the recent observations, non-equilibrium thermodynamics through particle creation mechanism may explain not only the late time acceleration but also exhibits early inflationary scenario, and, predict future transition to decelerating era again, and this transient phenomenon is supported by the works in \cite{Carvalho1, GL2011, SSS2014}. Finally, we remark that only future evolution of the universe can test whether our prediction from particle creation mechanism is correct, or, wrong.

\section{Acknowledgments}
SP acknowledges CSIR, Govt. of India for financial support through SRF scheme (File No: 09/096 (0749)/2012-EMR-I). SC thanks UGC-DRS programme at the Department of Mathematics, Jadavpur University. Both the authors thank Inter University Centre for Astronomy and Astrophysics (IUCAA), Pune, India for their warm hospitality as a part of the work was done during a visit there. We thank Prof. J.D. Barrow for introducing some references which were useful for the present work, and our other works. Finally, we are thankful to the anonymous referees for their valuable comments on the earlier version of the paper which helped us to improve the paper considerably.

\end{document}